\documentclass[final,5p,times,twocolumn]{elsarticle}
\usepackage{lineno,hyperref,amsmath,amsthm,amssymb,amsfonts,ragged2e,color}
\journal{``High Energy Density Physics''}
\begin{document}
\begin{frontmatter}
\title{Dust-acoustic rogue waves in four component plasmas}
\author{S. Jahan$^*$, A. Mannan, N.A. Chowdhury, and A.A. Mamun}
\address{Department of Physics, Jahangirnagar University, Savar, Dhaka-1342, Bangladesh\\
Email: $^*$jahan88phy@gmail.com}
\begin{abstract}
A theoretical investigation has been made on modulational instability (MI) and dust-acoustic
(DA) rogue waves (DARWs)  in a four dusty plasma medium containing inertial negatively charged massive heavy (light)
cold (hot) dust grains as well as super-thermal electrons and non-thermal ions. The
reductive perturbation method is used to derive the nonlinear Schr\"{o}dinger equation,
and two types of modes, namely fast and slow DA modes, have been observed. The conditions for
the MI and the formation of associated DARWs are found to be significantly modified by
the effects of non-thermality of ions ($\alpha$), super-thermality of electrons ($\kappa$),
density-ratio of non-thermal ion to cold dust ($\mu_i$), and   mass-ratio of cold dust to
hot dust ($\sigma$), etc. The implications of our current investigation in space and
laboratory plasmas are briefly discussed.
\end{abstract}
\end{frontmatter}
\section{Introduction}
\label{2sec:Introduction}
Now-a-days, scientist are enthralled by the natural elegance of dusty plasmas (DP) because of their
existence in space (viz. Saturn F-ring \cite{Akhtar2007}, circum-solar dust
grains \cite{Selim2015}, and interstellar molecular clouds \cite{Shukla2002}, etc.)
and enormous application in laboratory (viz. fusion device, rf discharges, and
Q-machine \cite{Shukla2002}, etc.). DP can be defined as low temperature fully or
partially ionized conducting gases whose constituents are electrons, ions, charged dust grains,
and neutral atoms. Normally, the dust grains of the DP are highly charged and massive
object (billions times higher than the protons) and their size ranges from micrometer to millimeters
\cite{Shukla2002}). Dust grains acquire a negative charge by the collection of electrons and the
mechanisms of electrons collection are photo-electric emission \cite{Shukla2002,Feuerbacher1972},
plasma currents \cite{Shukla2002}, and other process, etc. Almost 30 years ago, Rao \textit{et al.} \cite{Rao1990} first theoretically
predicted the presence of extremely low phase velocity dust-acoustic (DA) waves (DAWs) in space DP in
which the dust grains mass provide the inertia and the thermal pressures of electrons and ions
provide the restoring force to generate DAWs. A few years later, Barkan \textit{et al.} \cite{Barkan1995}
have experimentally verified the DAWs in plasmas containing negatively charged dust grains.

The formation of the nonlinear electrostatic structures in a plasma medium is rigourously depended on the velocity
distribution functions of the plasma species. The Maxwellian velocity distribution is applicable for the plasma
medium in which the plasma species are thermally equilibrium. But, it is often observed that
the plasma species in space as well as laboratory
plasmas are not in thermally equilibrium, and their basic features may deviate from the Maxwellian distribution.
Sometimes these thermally non-equilibrium plasma species may be governed by the non-thermal \cite{Cairns1995,Mamun1996,Elwakil2010} or super-thermal distribution \cite{Chowdhury2017,Sultana2014}. The non-thermal distribution function was first
introduced by Cairns \textit{et al.} \cite{Cairns1995} and have shown that a plasma medium in presence of non-thermal
electrons and cold ions could support the simultaneous existence of solitons within both
positive and negative potential. Mamun \textit{et al.} \cite{Mamun1996} have considered a model composed
of non-thermal ions and negatively charged dust.

The research regarding on modulational instability (MI) of DAWs and associated
dust-acoustic rogue waves (DARWs), which are the result of  unusual disturbances, in
nonlinear and dispersive mediums has been  vigorously increasing day by day.
A number of authors have employed reductive perturbation method (RPM) to obtain the nonlinear
Schr\"{o}dinger equation (NLSE)  to study the MI of the DAWs as well as formation of DARWs
in dusty plasma medium. Elwakil \textit{et al.} \cite{Elwakil2010} have
examined the effects of the non-thermal electrons on the MI of ion-acoustic waves
and found that the non-thermality of the electrons
decreases the stable domain of the ion-acoustic waves.
Sultana \textit{et al.} \cite{Sultana2014} have investigated the
MI in four component DP in presence of super-thermal electrons.
Chowdhury \textit{et al.} \cite{Chowdhury2017} have
studied the DARWs in presence of the super-thermal or $\kappa$-distributed
electrons in space DP and observed that within the fast DA modes,
the amplitude and the width of the DARWs are decreasing with $\kappa$.
Selim \textit{et al.} \cite{Selim2015} have studied the nonlinear DARWs in
presence of two-temperature (hot and cold) highly charged dust grains
and reported that amplitude and width of the DARWs decrease with hot
dust charge state. Elwakil \textit{et al.} \cite{Elwakil2013} have
investigated the effect of non-thermality of ions on the nature of DAWs
in two temperatures highly charged dust grains. Akhter \textit{et al.} \cite{Akhtar2007}
have  studied the DA solitary waves in the presence of hot and cold dust grains
along with iso-thermal electron and ions, and observed that  the
charge to mass ratio of each dust component plays a vital role for
the formation of the solitary structure in DP.
To the best knowledge of the authors, no attempt has been to study the MI of the DAWs
and associated DARWs in four component space DP  medium containing negatively charged hot and cold dust
as well as super-thermal electrons and non-thermal ions. The aim of the present investigation
is therefore to extend the work of Akhter \textit{et al.} \cite{Akhtar2007} to investigate the conditions for the MI
of the DAWs as well as the formation of the DARWs in four-component space DP medium
(in which the inertia is provided by the masses of the cold and hot dust grains
and restoring force is provided by the thermal pressure of the super-thermal electrons and non-thermal ions).

The rest of the manuscript is organized in the following scheme: The model equations of our
considered plasma system are stated in Sec. \ref{2sec:Model Equations}. The stability of the DAWs and the formation of
rogue waves are examined in Sec. \ref{2sec:Modulational instability and rogue waves}. 
Finally,  a conclusion is provided in Sec. \ref{2sec:Conclusion}.
\section{Model Equations}
\label{2sec:Model Equations}
we consider a collisionless, fully ionized, unmagnetized four-component DP system composed
of super-thermal electrons (charge $-e$, mass $m_e$), non-thermal ions (charge $+e$; mass $m_i$) and
inertial cold negatively charged dust grains (charge $q_c=-Z_ce$, mass $m_c$) as well as hot dust
grains (charge $q_h=+Z_he$; mass $m_h$); where $Z_c$ ($Z_h$) is the charge state of the negatively
charged cold (hot) dust grains. The governing equation of our plasma model can be written as:
\begin{eqnarray}
&&\hspace*{-1.3cm}\frac{\partial N_c}{\partial T}+\frac{\partial}{\partial X}(N_c U_c)=0,
\label{2eq:1}\\
&&\hspace*{-1.3cm}\frac{\partial U_c}{\partial T} + U_c\frac{\partial U_c }{\partial X}=\frac{Z_c e}{m_c}
\frac{\partial \varphi}{\partial X},
\label{2eq:2}\\
&&\hspace*{-1.3cm}\frac{\partial N_h}{\partial T}+\frac{\partial}{\partial X}(N_h U_h)=0,
\label{2eq:3}\\
&&\hspace*{-1.3cm}\frac{\partial U_h}{\partial T} + U_h\frac{\partial U_h }{\partial X}=\frac{Z_h e}{m_h}
\frac{\partial \varphi}{\partial X}-\frac{1}{m_h N_h}\frac{\partial P_h}{\partial X},
\label{2eq:4}\
\end{eqnarray}
where $N_c$ ($N_h$) is the number densities of the negatively charged cold (hot) dust grains;
$T$ ($X$) is the time (space) variable; $U_c$ ($U_h$) is the fluid speed of the negatively charged
cold (hot) dust species; $e$ is the magnitude of the charge of the electron; $\varphi$ is the
electrostatic wave potential, and $P_h$ is the adiabatic pressure of the negatively charged hot dust
grains. The system is closed through Poisson's equation
\begin{eqnarray}
&&\hspace*{-1.3cm}\frac{\partial^2\varphi}{\partial X^2}=4\pi e (N_e-N_i+Z_c N_c+Z_h N_h),
\label{2eq:5}
\end{eqnarray}
where $N_i$ and $N_e$ are, respectively, the ion and electron number densities. Now, in terms of
normalized variables, namely, $n_c=N_c/n_{c0}$ (with $n_{c0}$ being the equilibrium number
densities of the negatively charged cold dust grains), $n_h=N_h/n_{h0}$ (with $n_{h0}$ being
the equilibrium number densities of the negatively charged hot dust grains), $u_c=U_c/C_{dc}$
(with $C_{dc}$ being the sound speed of the negatively charged cold dust grains),
$C_{dc}=(Z_ck_BT_i/m_c)^{1/2}$ (with $T_i$ being the temperature of non-thermal ion and $k_B$ is the Boltzman constant), $u_h=U_h/C_{dc}$;
$\phi=e\varphi/k_BT_i$; $t=T \omega_{pdc}$ (with $\omega_{pdc}$ being the plasma frequency of the negatively
charged cold dust grains), $\omega_{pdc}=(4\pi e^2Z_c^2
n_{c0}/m_c)^{1/2}$, $x=X/\lambda_{Ddc}$ (with $\lambda_{Ddc}$ being the Debye length of
the negatively charged cold dust species), $\lambda_{Ddc}=(k_BT_i/4 \pi e^2 Z_c n_{c0})^{1/2}$;  $P_h=P_{h0}
(N_h/n_{h0})^\gamma$ [with being $P_h$ ($P_{h0}$) are the adiabatic (equilibrium adiabatic)
pressure of the negatively charged hot dust grains and $\gamma=(N+2)/N$, where $N$ is the degree of freedom, for one-dimensional
case, $N=1$ so that $\gamma=3$], and $P_{h0}=n_{h0}k_BT_h$ (with $T_h$ being the temperature of the negatively charged
hot dust grains). It may be noted here that for numerical analysis we have considered
that $m_c>m_h$, $Z_c>Z_h$, $n_{c0}>n_{h0}$, and $T_i, T_e>>T_h$ (with $T_e$ being the temperature of the
super-thermal electron). By normalization, \eqref{2eq:1}-\eqref{2eq:5}
becomes
\begin{eqnarray}
&&\hspace*{-1.3cm}\frac{\partial n_c}{\partial t}+\frac{\partial}{\partial x}(n_c u_c)=0,
\label{2eq:6}\\
&&\hspace*{-1.3cm}\frac{\partial u_c}{\partial t} + u_c\frac{\partial u_c}{\partial x}=
\frac{\partial \phi}{\partial x},
\label{2eq:7}\\
&&\hspace*{-1.3cm}\frac{\partial n_h}{\partial t}+\frac{\partial}{\partial x}(n_h u_h)=0,
\label{2eq:8}\\
&&\hspace*{-1.3cm}\frac{\partial u_h}{\partial t} + u_h\frac{\partial u_h}{\partial x}+
\delta n_h\frac{\partial n_h}{\partial x}=\sigma \frac{\partial \phi}{\partial x},
\label{2eq:9}\\
&&\hspace*{-1.3cm}\frac{\partial^2 \phi}{\partial x^2}=(\mu_i-\lambda-1)n_e-\mu_i n_i+n_c
+\lambda n_h,
\label{2eq:10}
\end{eqnarray}
where $\delta=3 T_h m_c/Z_c T_i m_h$, $\sigma=Z_h m_c/Z_c m_h$, $\mu_i=n_{i0}/Z_c n_{c0}$ (with $n_{i0}$
being the equilibrium number densities of the non-thermal ions), and $\lambda=Z_h n_{h0}/Z_c n_{c0}$.
The quasi-neutrality condition at equilibrium can be written as
\begin{eqnarray}
&&\hspace*{-1.3cm}n_{i0}=n_{e0}+Z_c n_{c0}+Z_h n_{h0},
\label{2eq:11}
\end{eqnarray}
where $n_{e0}$ is the equilibrium number densities of the super-thermal electrons. The number densities of
the super-thermal electrons and  non-thermal ions \cite{Cairns1995} can be written by the following normalized equation
\begin{eqnarray}
&&\hspace*{-1.3cm}n_e=\Big[1-\frac{\eta \phi}{(\kappa-3/2)} \Big]^{-\kappa+1/2},
\label{2eq:12}\\
&&\hspace*{-1.3cm}n_i=(1+\beta \phi+\beta \phi^2) ~\mbox{exp}(-\phi),
\label{2eq:13}
\end{eqnarray}
where $\kappa$ is the super-thermality of electrons and $\beta=4\alpha/(1+3\alpha)$ (with $\alpha$ being the non-thermality of ions)
and $\eta=T_i/T_e$ ($T_e>T_i$). Now, by substituting \eqref{2eq:11} and \eqref{2eq:12}
into \eqref{2eq:10}, and extending up to third order in $\phi$, one
can obtain the following relation
\begin{eqnarray}
&&\hspace*{-1.3cm}\frac{\partial^2 \phi}{\partial x^2}=n_c + \lambda n_h-\lambda-1+
\gamma_1 \phi+\gamma_2\phi^2+\gamma_3 \phi^3+\cdot\cdot\cdot,
\label{2eq:14}
\end{eqnarray}
where
\begin{eqnarray}
&&\hspace*{-1.3cm}\gamma_1=\frac{[\eta(\mu_i-\lambda-1)(\kappa-1/2)-\mu_i(\beta-1)
(\kappa-3/2)]}{(\kappa-3/2)},
\nonumber\\
&&\hspace*{-1.3cm}\gamma_2=\frac{[\eta^2(\mu_i-\lambda-1)(\kappa-1/2)(\kappa+1/2)
-\mu_i(\kappa-3/2)^2]}{2(\kappa-3/2)^2},
\nonumber\\
&&\hspace*{-1.3cm}\gamma_3=\frac{[\eta^3(\mu_i-\lambda-1)(\kappa-1/2)(\kappa+1/2)
(\kappa-3/2)+\prod]}{6(\kappa-3/2)^3},
\nonumber\
\end{eqnarray}
where $\prod=3\mu_i(\beta+3)(\kappa-3/2)^3$.
We will employ the RPM to derive the NLSE to study the MI of the DAWs.
The independent variables are stretched as $\xi=\epsilon (x-v_gt)$ and $\tau=\epsilon^2t$, where
$\epsilon$ is a small expansion parameter and $v_g$ is the group velocity of the IAWs.
The dependent variables \cite{Chowdhury2018} can be expressed as
\begin{eqnarray}
&&\hspace*{-1.3cm}\Upsilon(x,t)=\Upsilon_0+\sum_{m=1}^{\infty}\epsilon^{(m)}\sum_{l=-\infty}^
{\infty}\Upsilon_l^{(m)}(\xi,\tau) ~\mbox{exp}[il\Lambda],
\label{2eq:15}
\end{eqnarray}
where $\Upsilon_{l}^{(m)}=[n_{cl}^{(m)},u_{cl}^{(m)},n_{hl}^{(m)},u_{hl}^{(m)},\phi_l^{(m)}]$,
$\Lambda_0=[1,0,1,0,0]^\tau$, $\Lambda=(kx-\omega t)$, and $k$ ($\omega$) can be represented
as the carrier wave number (frequency), respectively. We are following parallel
mathematical steps as Chowdhury \textit{et al.} \cite{Chowdhury2018} have done in their work to obtain successively
the IAWs dispersion relation, group velocity, and NLSE. 
The dispersion relation for the DAWs can be expressed as
\begin{eqnarray}
&&\hspace*{-1.3cm}\omega^2=\frac{B \pm \sqrt{B^2-4ME}}{2M},
\label{2eq:16}
\end{eqnarray}
where $B=\delta (k^2+\gamma_1)+\lambda \sigma+1$, $M=(k^2+\gamma_1)/k^2$, and $E=\delta k^2$.
It may be noted that the real and positive values of $\omega$ can be obtained under the
satisfactory condition of $B^2>4ME$. One may recognize the fast ($\omega_f$)/slow ($\omega_s$)
DA modes can according to to the positive/negative sign of \eqref{2eq:18}.
The group velocity $v_g$ of DAWs can be written as 
\begin{eqnarray}
&&\hspace*{-1.3cm}v_g=\frac{\lambda \sigma \omega^5+\delta \lambda \sigma k^2 \omega^3+2 \omega A^2
-\lambda \sigma A \omega^3-2 A^2 \omega^3}{2 k (A^2+\lambda \sigma \omega^4)}.
\label{2eq:17}
\end{eqnarray}
where $A=\delta k^2-\omega^2$. Finally, the NLSE can be written as:
\begin{eqnarray}
&&\hspace*{-1.3cm}i\frac{\partial \Phi}{\partial \tau}+P\frac{\partial^2 \Phi}{\partial \xi^2}
+Q|\Phi|^2\Phi=0,
\label{2eq:18}
\end{eqnarray}
where $\Phi=\phi_1^{(1)}$ for simplicity and $P$ ($Q$) is the dispersion (nonlinear) coefficient,
and is written by
\begin{eqnarray}
&&\hspace*{-1.3cm}P=\frac{C_1}{2 \omega A k^2 (A^2+\lambda \sigma \omega^4)}
\nonumber\\
&&\hspace*{-1.3cm}Q=\frac{C_2}{2 k^2 (A^2+\lambda \sigma \omega^4)}
\nonumber\
\end{eqnarray}
\begin{figure}[t!]
\centering
\includegraphics[width=80mm]{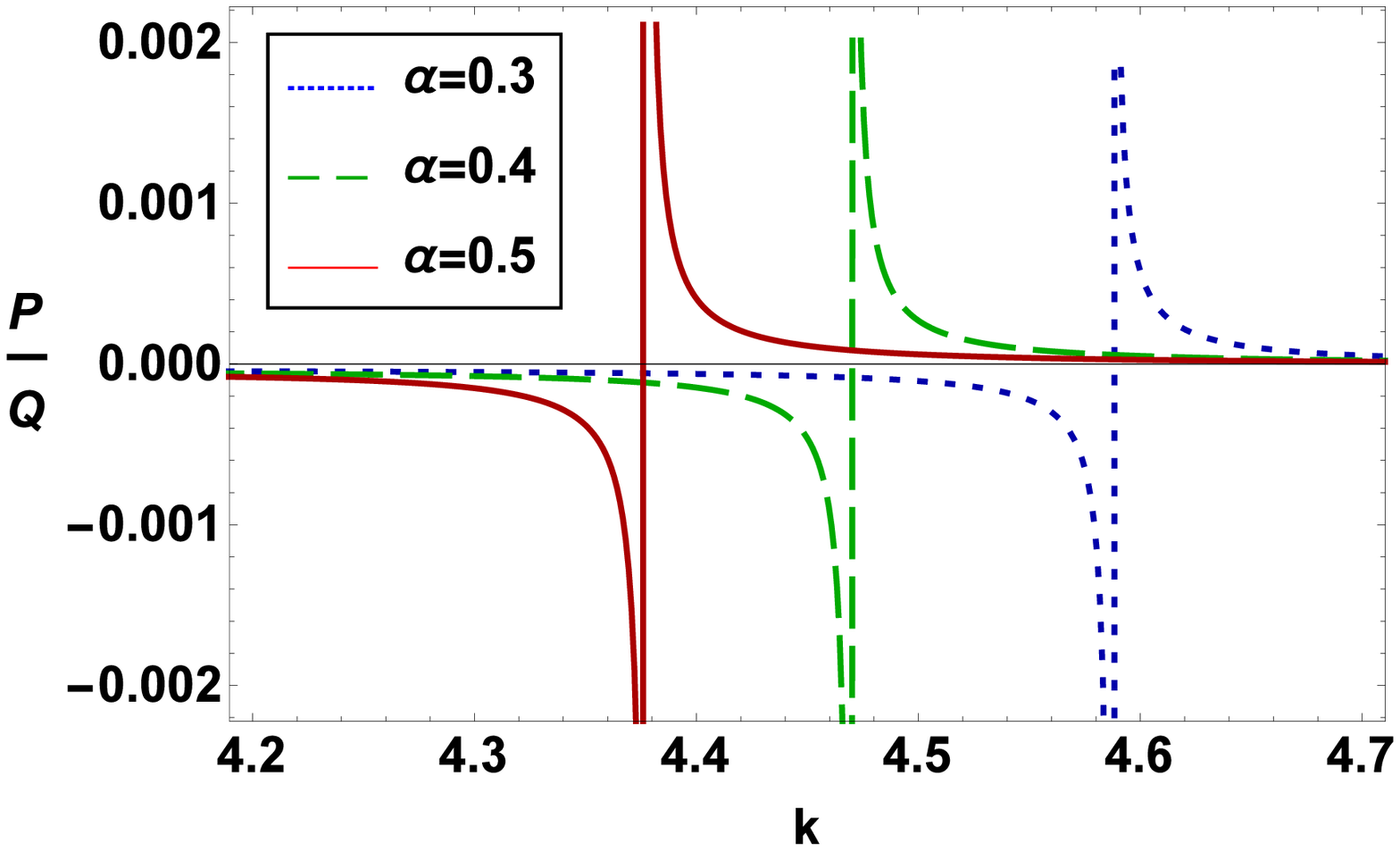}
\caption{The variation of $P/Q$ with $k$ for different values of $\alpha$;
along with  $\kappa=1.7$, $\delta=0.0006$, $\lambda=0.5$, $\mu_i=2.0$, $\sigma=1.6$, $\eta=0.8$, and $\omega_f$.}
\label{2Fig:F1}
\vspace{0.8cm}
\includegraphics[width=80mm]{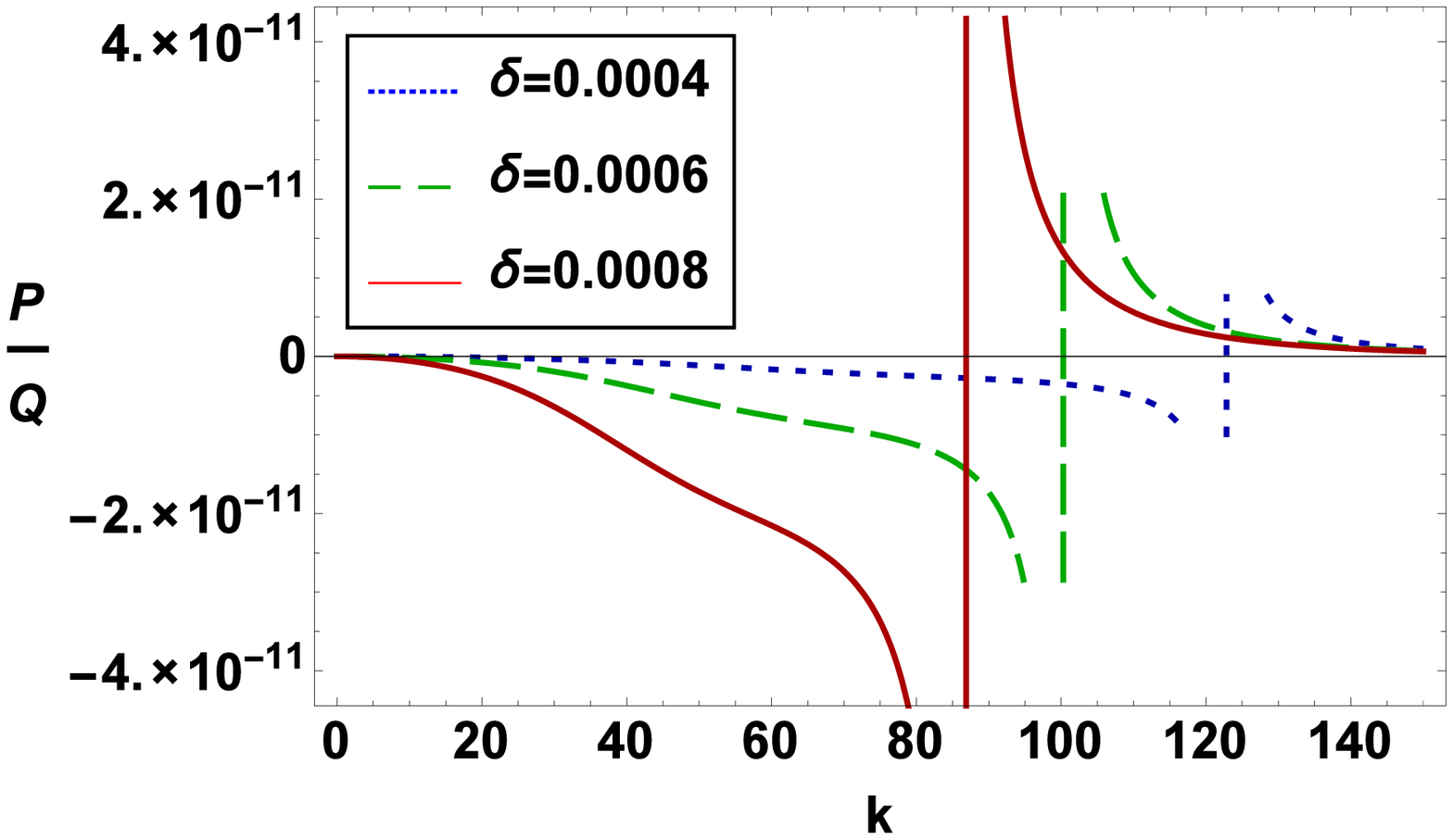}
\caption{The variation of $P/Q$ with $k$ for different values of $\delta$;
along with $\alpha=0.3$ $\kappa=1.7$, $\lambda=0.5$, $\mu_i=2.0$, $\sigma=1.6$, $\eta=0.8$, and $\omega_s$.}
\label{2Fig:F2}
\end{figure}
where
\begin{eqnarray}
&&\hspace*{-1.3cm}C_1=\lambda \sigma \omega^4 (\omega-k v_g) (2 k v_g \omega^2+A k v_g
-\omega^3-\delta \omega k^2)
\nonumber\\
&&\hspace*{-0.5cm}+\lambda \sigma k \omega^4 (\delta k-\omega v_g) (2 \omega k v_g+A
-\omega^2-\delta k^2)
\nonumber\\
&&\hspace*{-0.5cm}+A^3 (\omega-k v_g)(\omega-3 k v_g)-A^3 \omega^4,
\nonumber\\
&&\hspace*{-1.3cm}C_2=3 \gamma_3 A^2 \omega^3+2 \gamma_2 A^2 \omega^3 (C_7+C_{12})
-2 A^2 k^3 (C_4+C_9)
\nonumber\\
&&\hspace*{-0.5cm}-2 \lambda \sigma k^3 \omega^4 (C_6+C_{11})-\omega A^2 k^2 (C_3+C_8)
\nonumber\\
&&\hspace*{-0.5cm}-\sigma \lambda k^2 \omega^3(\omega^2+\delta k^2) (C_5+C_{10}),
\nonumber\
\end{eqnarray}
\begin{eqnarray}
&&\hspace*{-1.3cm}C_3=\frac{3 k^4-2 C_5 \omega^2 k^2}{2 \omega^4},
\nonumber\\
&&\hspace*{-1.3cm}C_4=\frac{C_1 \omega^4-k^4}{k \omega^3},
\nonumber\\
&&\hspace*{-1.3cm}C_5=\frac{2 \sigma C_5 A^2 k^2-\delta \sigma^2 k^6-3 \sigma^2 \omega^2 k^4}{2 A^3},
\nonumber\\
&&\hspace*{-1.3cm}C_6=\frac{\omega C_3 A^2-\omega \sigma^2 k^4}{k A^2},
\nonumber\\
&&\hspace*{-1.3cm}C_7=\frac{3 A^3 k^4+2 \gamma_2 A^3 \omega^4-3 \lambda \sigma^2 k^4 \omega^6
-\delta \lambda
\sigma^2 \omega^4 k^6}{2 \omega^2 A^2 (A k^2-4 A \omega^2 k^2-\lambda \sigma \omega^2 k^2-
\gamma_1 A \omega^2)},
\nonumber\\
&&\hspace*{-1.3cm}C_8=\frac{2 v_g k^3+\omega k^2-C_{10} \omega^3}{v_g^2 \omega^3},
\nonumber\\
&&\hspace*{-1.3cm}C_9=\frac{C_6 v_g \omega^3-2 k^3}{\omega^3},
\nonumber\\
&&\hspace*{-1.3cm}C_{10}=\frac{2 \omega v_g \sigma^2 k^3+\sigma^2 \omega^2 k^2+
\delta \sigma^2 k^4-\sigma C_{10} A^2}{A^2 (v_g^2-\delta)},
\nonumber\\
&&\hspace*{-1.3cm}C_{11}=\frac{C_8 v_g A^2-2 \omega \sigma^2 k^3}{A^2},
\nonumber\\
&&\hspace*{-1.3cm}C_{12}=\frac{C_{13}}{A^2 \omega^3 \left \{(v_g^2-\delta)+\lambda \sigma v_g^2
-\gamma_1 v_g^2 (v_g^2-\delta)\right \}},
\nonumber\\
&&\hspace*{-1.3cm}C_{13}=2 \gamma_2 A^2 v_g^2 \omega^3 (v_g^2-\delta)+A^2 (v_g^2-\delta)
(2 v_g k^3+\omega k^2)
\nonumber\\
&&\hspace*{-0.5cm}+\lambda v_g^2 \omega^3 (2 \omega v_g \sigma^2 k^3+\delta \sigma^2 k^4
+\sigma^2 \omega^2 k^2).
\nonumber\
\end{eqnarray}
\section{MI and rogue waves}
\label{2sec:Modulational instability and rogue waves}
The stability of the DAWs profile, which amplitude evolution is governed by the NLSE, is totally depended
on the the nonlinear ($P$) and dispersive ($Q$) co-efficient \cite{Chowdhury2018,Sultana2011,Rahman,Chowdhury,Kourakis2006,Fedele2002a}.
It should be noted that the nonlinear and the dispersive co-efficient can be either positive or negative
according to the values of $k$, $\alpha$, $\mu_i$, $\delta$, $\kappa$, and $\sigma$, etc. In our present
numerical analysis, for small wave number, i.e., $k\rightarrow 0$, the sign of $P$ is negative, but
for large wave number, i.e., $k\rightarrow 70$, the sign of $P$ is positive (figure is not included). Generally,
the sign of $P$ is always negative without consideration of degenerate or adiabatic
pressure in the momentum equation \cite{Sultana2011}, but if any one consider the degenerate/adiabatic pressure term
in their momentum equation, it may possible for $P$ to obtain positive or negative values
depending on the $k$ and other plasma parameters \cite{Kourakis2006}. On the other hand, for small wave number,
i.e., $k\rightarrow 0$, the values of $Q$ is positive, but for large wave number,
i.e., $k\rightarrow 5$, the values of $Q$ is negative.
\begin{figure}[t!]
\centering
\includegraphics[width=80mm]{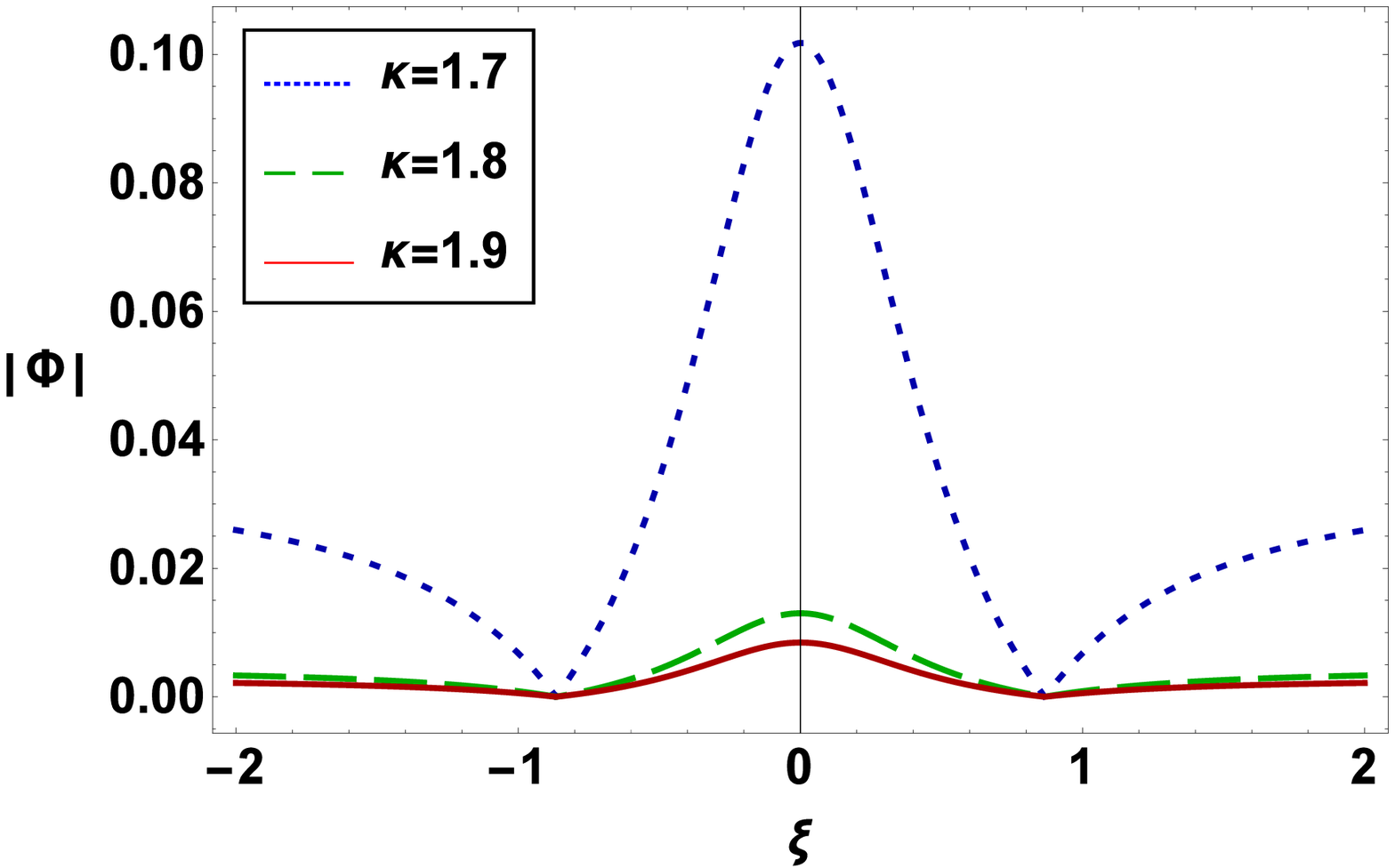}
\caption{The variation of $|\Phi|$ with $\xi$ for different values of $\kappa$;
along with $\alpha=0.3$ $\delta=0.0006$, $\lambda=0.5$, $\mu_i=2.0$, $\sigma=1.6$, $\eta=0.8$, and $\omega_f$.}
\label{2Fig:F3}
\vspace{0.8cm}
\includegraphics[width=80mm]{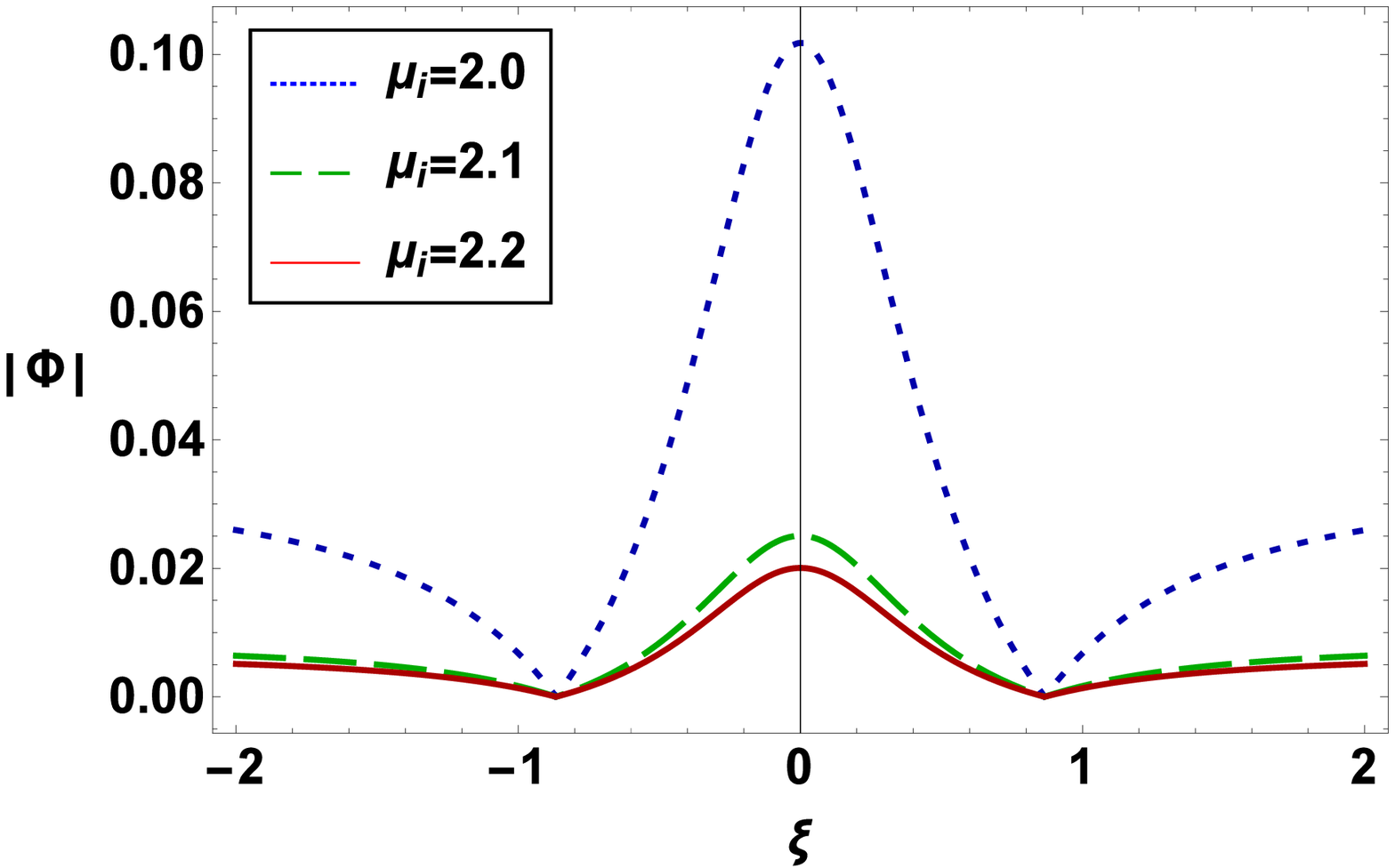}
\caption{The variation of $|\Phi|$ with $\xi$ for different values of $\mu_i$;
along with $\alpha=0.3$ $\delta=0.0006$, $\kappa=1.7$, $\lambda=0.5$, $\sigma=1.6$, $\eta=0.8$, and $\omega_f$.}
\label{2Fig:F4}
\vspace{0.8cm}
\includegraphics[width=80mm]{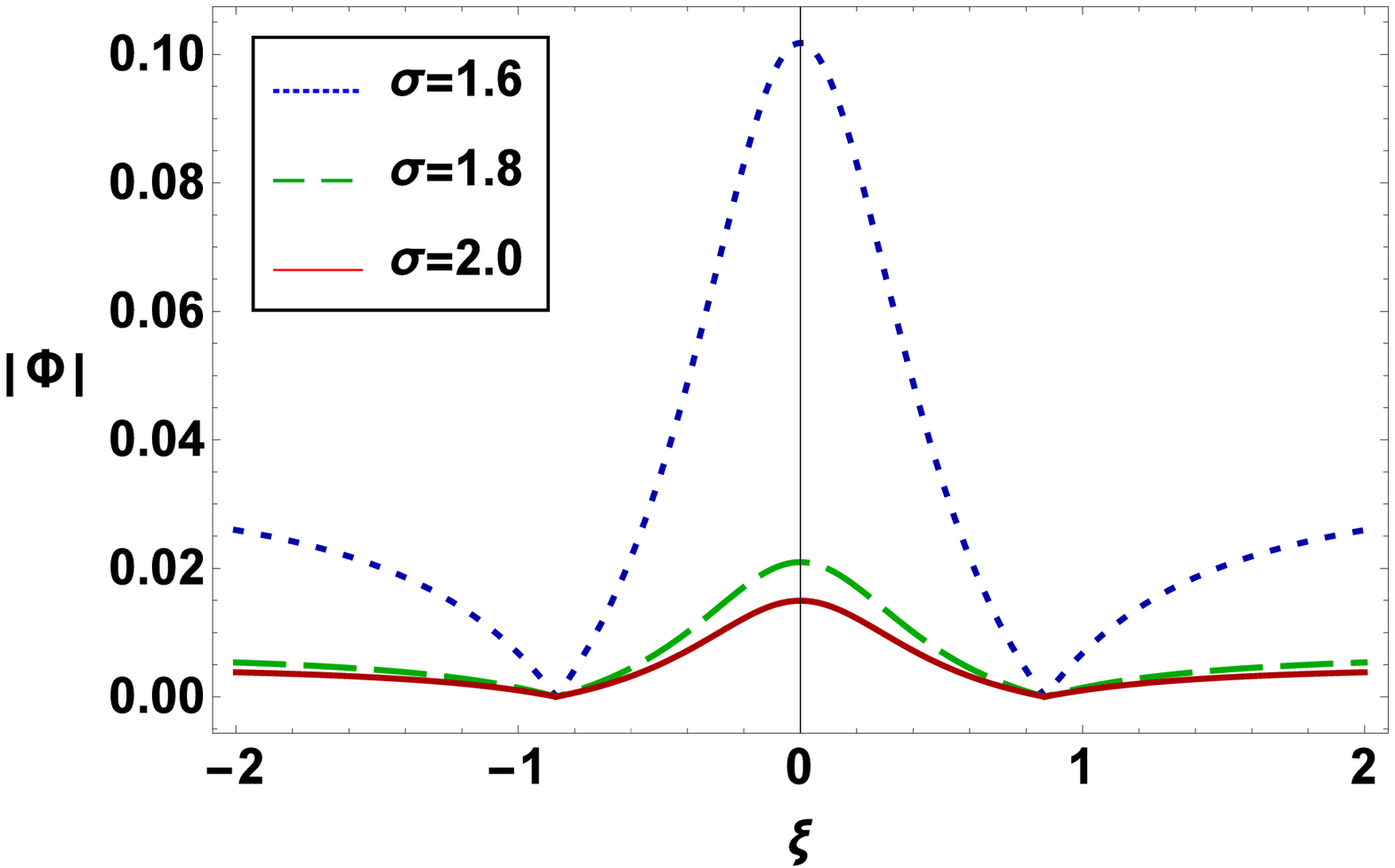}
\caption{The variation of $|\Phi|$ with $\xi$ for different values of $\sigma$;
along with $\alpha=0.3$ $\delta=0.0006$, $\kappa=1.7$, $\lambda=0.5$, $\mu_i=2.0$, $\eta=0.8$, and $\omega_f$.}
\label{2Fig:F5}
\end{figure}
The criteria of the MI of DAWs and the formation of the highly energetic nonlinear structures in four component DP medium
can be regulated by the sign of $P$ and $Q$. The DAWs are modulationally stable when $P/Q<0$ ($P$ and $Q$
have the different sign). On the other hand, when $P/Q>0$ ($P$ and $Q$ have the same sign)
the DAWs are modulationally unstable. The intersecting point, which can open a new unstable
window from the stable domain for DAWs in nonlinear and dispersive plasma medium, between $P/Q$ curve with
$k$-axis ($P/Q$ vs. $k$ graph) is referred to as the critical wave number ($k_c$). The nature of the $P/Q$ curve with
plasma parameter $\alpha$ and $\delta$ can be observed from Figs. \ref{2Fig:F1} and \ref{2Fig:F2},
and it is obvious from these two figures that (a) the modulationally stable and unstable domain for the DAWs in nonlinear and dispersive medium
can be found for the fast ($\omega_f$) and slow ($\omega_s$) DA modes according to the sign (positive) of equation \eqref{2eq:16};
(b) the value of $k_c$ decreases with non-thermality of the ions (via $\alpha$);
(c) so, excess non-thermality reduces the stable domain of the DAWs and this result 
is a good agreement with the result of Elwakil \textit{et al.} \cite{Elwakil2010} work;
(d) the stable region of DAWs increases (decreases) with ion  (hot dust) temperature for
constant value of  cold and hot dust mass, and  hot dust charge state (via $\delta$).

The first-order rational rogue wave  solution (developed by Darboux Transformation Scheme) in the
unstable region ($P/Q>0$) can be written as \cite{Anikiewicz2009}
\begin{eqnarray}
&&\hspace*{-1.0cm}\Phi(\xi,\tau)=\sqrt{\frac{2P}{Q}} \left[\frac{4(1+4iP\tau)}{1+
16P^2\tau^2+4\xi^2}-1 \right]\mbox{exp}(2iP\tau).
\label{2eq:19}
\end{eqnarray}
The rogue wave solution inform us that the wave-particle interaction occurs within a small evolution of time and space.
The effects of different plasma parameters in the formation of the DARWs can be 
observed from Figs. \ref{2Fig:F3}-\ref{2Fig:F5}.
From Fig. \ref{2Fig:F3}, it can be observed that the amplitude of the DARWs decreases with the increase in the value  $\kappa$.
From Literature, it can be manifested that small value of $\kappa$ represents a  hard spectrum, and when $\kappa$ is small
then the movement of the plasma particles is larger then the large value of $\kappa$. So, the nonlinearity of the
plasma medium increases with the decrease of $\kappa$, and this result agrees with the result of Chowdhury \textit{et al.} \cite{Chowdhury2017} work.

The nature of the amplitude and the width of the DARWs is also governed by plasma parameters
$\mu_i$ and $\sigma$, and their effects can be observed from Figs. \ref{2Fig:F4} and \ref{2Fig:F5}.
It is obvious from Figs. \ref{2Fig:F4} and \ref{2Fig:F5} that (a) the height and the width of the
DARWs decrease with the increase in the value of equilibrium ion number density ($n_{i0}$)
for fixed values of $Z_c$ and $n_{c0}$ (via $\mu_i$); (b) the nonlinearity of the plasma
medium increases (decreases)  with $m_h$ ($m_c$) by depicting taller (smaller) DARWs for fixed values of $Z_c$ and $Z_h$ (via $\sigma$);
(c) on the other hand, the nonlinearity of the plasma medium increases with the increase in the value of hot dust charge state $Z_c$,
decreases with increase of the cold dust charge state $Z_h$ when their hot and cold dust mass remain constant. So, the
mass and charge state of the hot and cold dust grains provide an opposite effects in the plasma medium to generate highly energetic DARWs.
\section{Conclusion}
\label{2sec:Conclusion}
In this work, we have considered an unmagnetized four-component DP system consisting of
negatively charged cold and hot dust grains, electrons following super-thermal $\kappa$-distribution and
ions following non-thermal distribution. The standard NLSE is used to recognize the
stable and unstable domain for the DAWs according to the sign of nonlinear and dispersive coefficients.
The results of our present investigation will be useful in understanding the propagation of DAWs, MI of DARWs, and
finally the generation of the DARWs in space plasma medium (viz. Saturn F-ring \cite{Akhtar2007}, circum-solar dust
grains \cite{Selim2015}, and interstellar molecular clouds \cite{Shukla2002}, etc.) and also
laboratory DP (viz. fusion device, rf discharges, and Q-machine \cite{Shukla2002}, etc.).

\end{document}